# Transient growth and evaporation kinetics at vicinal surfaces with non-local electromigration: step bunching instability

B. Ranguelov, S. Stoyanov
*Institute of Physical Chemistry, Bulgarian Academy of Sciences, 1113,Sofia, Bulgaria*

**Abstract**
The steps at the crystal surfaces could be transparent for the migrating adatoms. In the case of significant transparency the velocity of a given step in a given moment is affected by detachment of atoms from rather distant steps in rather earlier moments. The reason is that the adatom needs time to cross many steps before attaching to a kink position. The theory of the crystallization should account for the time dependence of the adatom concentrations on the terraces. Such a transient growth model is developed here for the limit of slow kinetics at the steps and fast surface diffusion. The model accounts for the non-local electromigration of the adatoms. The central results of this study are:

1. The vicinal surface with transparent steps is unstable when the drift velocity $V_{drift}$ of the adatoms multiplied with the relative deviation of the adatom concentration from equilibrium $\frac{n_s^e - n_s}{n_s^e}$ is larger than the critical velocity of the steps $V_{cr}$. It is important to note that step bunching instability takes place at a positive value of the product $\frac{n_s^e - n_s}{n_s^e} V_{drift}$. Since $\frac{n_s^e - n_s}{n_s^e}$ is positive during sublimation and negative during growth step bunching takes place at a positive drift velocity in the first case and at a negative $V_{drift}$ (step-down direction of the electromigration) in the second case.

2. The most unstable mode has a wavelength (expressed as a number of terraces) given by the expression $\lambda_{\max} = \pi\sqrt{2P_k}$. In other words, the wavelength (as a number of terraces) is independent of the surface orientation (the miscut angle of the vicinal) as well as on the magnitude of step repulsion and the electromigration force. The quantity $\lambda_{\max}$ depends only on the relative transparency of the steps $P_k = P/K$, where $P$ and $K$ are the transparency coefficient and the step kinetic coefficient respectively.

## 1. Introduction
The most intriguing feature of the step bunching, induced by the direct electric current flowing through the Si wafer, is the complicated temperature dependence of this phenomenon [1-5]. The experiments show step bunching to occur at different directions of the electric current when the temperature of the Si crystal increases from $860°C$ to $1350°C$. Homma and Aizawa [5] observed step bunching at step-down direction of the electric current in temperature intervals $860°C - 960°C$ and $1200°C - 1300°C$, whereas at $1060°C - 1200°C$ and above $1320°C$ step bunching takes place at step-up direction of the current (the temperatures reported by different authors vary at most by $50°C$ [1, 3, 4]) . To find an explanation of these experimental results is a challenging task especially after Yagi et al [6] established the electromigration of adatoms does not change its direction in the temperature interval $900°C - 1250°C$.

Step bunching at step-down direction of the electromigration of adatoms was satisfactorily explained by generalizing the Burton Cabrera Frank (BCF) model [7-9] to account for the drift of the adatoms [10, 11]. For the step bunching at step-up direction of the electromigration Stoyanov [12, 13] proposed a mechanism based on the transparency of the steps (an adatom has a considerable probability to migrate from a given terrace to the neighboring one without visiting a kink at the step separating the two terraces). On this basis he was able to prove that a moving bunch of steps is stable with step-up drift of the adatoms. He also studied the self-similarity of the shape of the bunch [13]. Later Metois and Stoyanov [14] predicted the step bunches (at step-up current) to disappear when the sublimation is replaced by growth of the silicon crystal. Step bunching reappears during growth provided the direction of the electric current is inversed from step-up to step-down. Experimental proof [14] of these predictions was an essential step to the understanding of the crystal surface instabilities. Linear stability analysis of the vicinal surface in the model proposed by Stoyanov [13] was published by Sato et al [15]. These authors studied also the time evolution of the bunching instability of transparent steps by performing Monte Carlo simulations of dynamics of steps on a square lattice. A discrete model of a vicinal crystal surface with transparent steps was formulated recently by Pierre-Louis [16]. He also treated in detail the linear stability of the transparent steps dynamics. Finally, let us mention that a rather different idea was advanced by Weeks et al [17] for the explanation of the step bunching at step-up direction of the adatom drift – the surface diffusion in the vicinity of a step is assumed to be faster than the diffusion in the central part of the terrace.

The present paper addresses the problem of electromigration induced instability of a vicinal surface by analyzing a discrete model with transparent steps. Our philosophy is to approach this phenomenon by non-steady state treatment similar to the considerations in our recent papers [18,19]. We think such a treatment is strongly justified in the case of significant transparency. The point is that the velocity of a given step in a given moment is affected by processes (detachment of atoms) that already took place at rather distant steps in rather earlier moments. The detached adatom need time to travel across say 10 terraces to finally attach to a kink position. To explore the impact of these features of the growth kinetics we go beyond the quasi-static approximation [7-9] in the calculation of the adatom concentration on the terraces of the crystal face and analyze the transient effects (like in our previous works [18, 19]).

The paper has the following structure – the next section introduces the basic concepts in the kinetics of crystal growth and sublimation in a presence of non-local electromigration. Section 3 is devoted to the formulation of the model and some comments on the stability of a train of transparent steps. These comments are based on an expression for the macroscopic mass current at the vicinal surface. Section 4 presents the derivation of kinetic equations, describing the time evolution of the terraces and the adatom concentrations on them. Section 5 contains the results of the linear stability analysis of the vicinal surface during sublimation and growth. Section 6 is devoted to the non-linear dynamics of steps, studied by numerical integration of the equations of step motion. The central results of the paper are summarized in Section 7.

**2. Basic concepts**
Transparency of the steps is defined through the boundary conditions of the differential equation describing the diffusion of the atoms adsorbed on the terraces of a vicinal crystal surface. In the case we analyze here (existence of a force $F$ applied on the adatoms) the boundary conditions read

$$D_s\left(-\frac{dn_i}{dx}+\frac{Fn_i}{kT}\right)=-K[n_i(x_i)-n_s^e(x_i)]+P[n_{i-1}(x_i)-n_i(x_i)] \qquad (1a)$$

at $x = x_i$ and

$$D_s\left(-\frac{dn_i}{dx}+\frac{Fn_i}{kT}\right)=K[n_i(x_{i+1})-n_s^e(x_{i+1})]+P[n_i(x_{i+1})-n_{i+1}(x_{i+1})] \qquad (1b)$$

at $x = x_{i+1}$. Here $D_s$ is the coefficient of surface diffusion and $n_i$ is the concentration of mobile atoms, adsorbed on the $i-th$ terrace (see Fig.1).

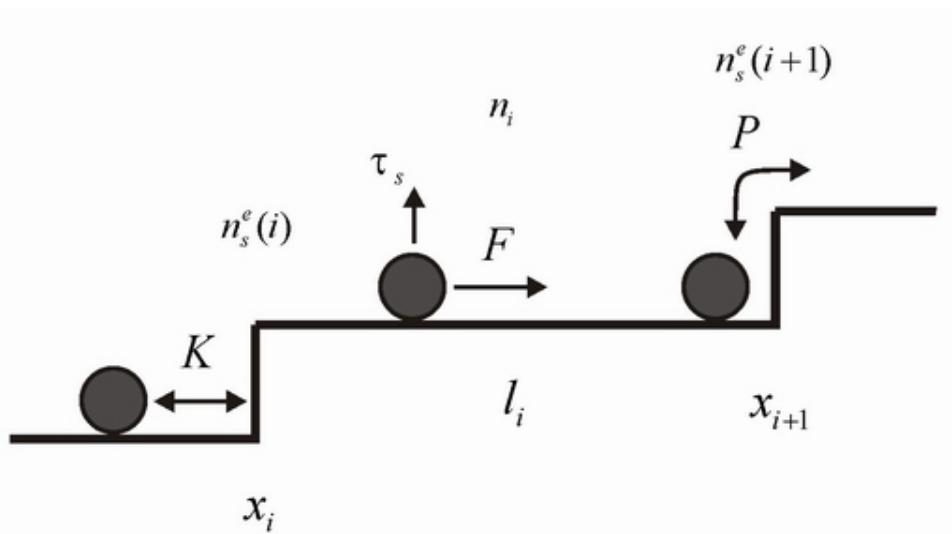

**FIG.1.** Model of the vicinal surface and adatoms: $F$ - electromigration force, $K$ - step kinetic coefficient, $P$ - transparency coefficient.

The terms in the right hand side of these equations have a clear physical meaning - the first term describes the exchange of atoms between the crystal phase and the two-dimensional dilute layer of adsorbed atoms. This exchange is characterised by the step kinetic coefficient $K$. The second term describes the inter-terrace exchange. The coefficient $P$ characterises the transparency of the steps - $P = 0$ means zero transparency, i.e. no atoms can cross the step. In other words all atoms attach to the step edge, migrate along it and eventually attach to a kink position. The attachment is not "for ever". As time proceeds the same atom can detach from a kink position and find itself again in a state of mobile adsorption. This is the mechanism the exchange of atoms between the crystal phase and the two-dimensional dilute layer of adsorbed atoms is realized. Non-zero transparency $P > 0$ means that the adatoms have a chance to either jump directly from one terrace to the neighbouring one, or effectively cross the step by attachment to the step edge and subsequent detachment from it before visiting a kink position. The boundary conditions (1a, 1b), accounting for both crystal – adlayer exchange and the inter-terrace exchange, were first formulated in Ref. [20] for the case $F = 0$, i.e. in absence of electromigration.

The coefficients $K$ and $P$ have a dimension of velocity but only $K$ is directly related to the crystallization kinetics and the motion of the elementary steps on the crystal surface through the equation

$$v(x_i) = \frac{dx_i}{dt} = -K\Omega[n_i(x_i) - n_s^e(x_i)] - K\Omega[n_{i-1}(x_i) - n_s^e(x_i)] \qquad (2)$$

Here $\Omega$ is the area of one atomic site at the crystal surface and the sign minus provides for a positive velocity of the steps during sublimation (see Fig.1) when $n_i(x_i) < n_s^e(x_i)$. The transparency coefficient $P$ has an impact on the gradient of the concentration of adatoms on the terraces. In this way $P$ could be essential for the stability of the step train with respect to the fluctuations of the step density.

On a simple physical ground one can assume that $P$ contains the same frequency factor like the surface diffusion coefficient $D_s$ and also a Boltzmann factor, i.e., $P = a\omega \exp(-E_{st}/kT)$ where $E_{st}$ is the activation energy for a transition from one terrace to the neighbouring one. Thus we can write for the characteristic length related to the step transparency

$$d_p = \frac{D_s}{P} = a\exp\left(\frac{E_{st} - E_{sd}}{kT}\right) \qquad (3)$$

where $a$ is the interatomic distance.

Since the condition for a considerable transparency is $d_p \ll l$ we obtain from (3) that the energy difference $E_{st} - E_{sd}$ should be of the order of $kT$.

Experimental evidences for considerable step transparency at Si(001) were reported by Tanaka et al [21].

### 3. Growth model – microscopic and macroscopic view

The external force $F$ acting on the adatoms induces a gradient of their concentration on a given terrace. Intuitively, one expects the step transparency to decrease the concentration gradient. To evaluate the gradient we analyze the simplest case of crystal annealing with negligible desorption and zero deposition rate. Then the surface transport on each terrace is described by the equation [10]

$$D_s \frac{d^2 n_i}{dx^2} - \frac{D_s F}{kT} \frac{dn_i}{dx} = 0 \qquad (4)$$

with $x$ being an axis, perpendicular to the step edges (a system of equidistant parallel steps with straight edges is under consideration).

The solution of the eq.(4) with boundary conditions (1a, 1b) reads

$$n_i(x) = \frac{n_s^e \left[\exp\left(\frac{Fx}{kT}\right) + \frac{l}{2d_s} + \frac{l}{2d_p}\right]}{1 + \frac{l}{2d_s} + \frac{l}{2d_p}} \qquad (5)$$

where $d_s = \frac{D_s}{K}$ and $d_p = \frac{D_s}{P}$ are characteristic lengths for step kinetics and step transparency, whereas $l$ is the terrace width. For the concentration gradient at $x=0$ the eq.(5) gives

$$grad(n_i) = \frac{n_s^e \frac{F}{kT}}{1 + \frac{l}{2d_s} + \frac{l}{2d_p}} \qquad (6)$$

In the next section we shall use this expression to construct a discrete model for the dynamics of transparent steps. Now, however, we shall briefly discuss the stability of the crystal surface on the basis of macroscopic concepts.

The electromigration force generates a macroscopic flux on the vicinal surface. This flux is easy to estimate by calculating the net number of atoms crossing (per unit time) a unit length of a given step in the direction of the force $F$. Since we focus our attention on the dynamics of strongly transparent steps (where the adatoms have much higher probability to jump to the next terrace than to visit a kink at the step, i.e., $d_s \approx l$ and $d_p \ll l$) the expression (6) can be simplified to

$$grad(n_i) \approx n_s^e \frac{2Fd_p}{kTl} \qquad (7)$$

This expression for the concentration gradient would be valid on the whole terrace when the mean diffusion distance is much larger than the terrace width ($\lambda_s = \sqrt{D_s \tau_s} \gg l$). Assuming the last inequality to be fulfilled we can write

$$n_s\left(\frac{l}{2}\right) = n_s^e\left(1 + \frac{Fd_p}{kT}\right) \qquad (8)$$

and now the macroscopic flux on a vicinal surface with a terrace width $l$ is

$$J_{mac} = P\left[n_{i-1}\left(\frac{l}{2}\right) - n_i\left(-\frac{l}{2}\right)\right] = 2Pn_s^e \frac{Fd_p}{kT} = n_s^e \frac{2FD_s}{kT} \qquad (9)$$

As seen the macroscopic flux does not depend on the orientation (the slope $m_x = \frac{h_0}{l}$) of the vicinal surface and the instability condition $\frac{dJ_{mac}}{dm_x} > 0$ (see [22, 23]) is not fulfilled. Therefore, as far as the annealing under negligible desorption and zero deposition rate (i.e., equilibrium conditions at the crystal surface) is concerned, no instability is expected. The situation, however, becomes different in a presence of considerable desorption or deposition rate.

Under these conditions the equilibrium concentration $n_s^e$ in eq.(9) should be replaced by the average (non-equilibrium) concentration $n_i$ determined from the steady state solution of the equation

$$\frac{dn_i}{dt} = R - \frac{n_i}{\tau_s} - \frac{2K}{l} n_i + \frac{2K}{l} n_s^e \qquad (10)$$

Thus one obtains

$$n_s = \frac{R\tau_s + \kappa n_s^e m_x}{1 + \kappa m_x} \qquad (11)$$

where $\kappa = 2K\tau_s / h_0$ and the instability criterion takes the form

$$\frac{dJ_{mac}}{dm_x} = 2P \frac{Fd_p}{kT} \frac{\kappa(n_s^e - R\tau_s)}{(1 + \kappa m_x)^2} > 0 \qquad (12)$$

It is essential to note that the eq.(12) definitely show the stability of the vicinal surface to depend on the sign of the product $F(n_s^e - R\tau_s)$. In other words, instability during sublimation ($(n_s^e - R\tau_s) > 0$) and growth ($(n_s^e - R\tau_s) < 0$) is expected to occur at step-up and step-down directions of the electromigration force (see Fig.1) in agreement with the experiment [4].

**4. Kinetic equations for the discrete model**

Now we start a detailed treatment of the dynamics of transparent steps. First we assume (in analogy with eq.(7)) that the gradient of the adatom concentration on the $i-th$ terrace is $grad(n_i) = n_i \frac{2Fd_p}{kTl}$ where $n_i$ is the average concentration on this terrace. The second assumption is that the gradient has approximately a constant value over the whole terrace (this is fulfilled when $\lambda_s \gg l$). The use of this expression for the adatom concentration gradient requires some comments. The expression (6) is obtained for a system of equidistant parallel steps with straight edges. For this configuration the expression (6) is an exact result. It could be a reasonable approximation when the step configuration slightly deviates from the equidistant one (as in the linear stability analysis given in Section 5). In the later stages of the step bunching instability, however, the terrace distribution is far from being uniform (neighbouring terraces could have rather different size). That is why the expression (6) is no more a good approximation for the adatom concentration gradient on the terraces at the vicinal crystal surface. Therefore the late stages of step bunching instability require special considerations which are beyond the scope of this paper. Here we shall make use of the expression (7) and write a differential equation for the time evolution of the concentration of adatoms on the $i-th$ terrace during sublimation. In the framework of this approximation we obtain

$$\frac{dn_i}{dt} = -\frac{n_i}{\tau_s} - \frac{2K}{l_i} n_i + \frac{K}{l_i}\left[n_s^e(i+1) + n_s^e(i)\right] - \frac{2P}{l_i} n_i + \frac{P}{l_i} n_{i-1}\left(1 + \frac{Fd_p}{kT}\right) + \frac{P}{l_i} n_{i+1}\left(1 - \frac{Fd_p}{kT}\right) \quad (13)$$

The terms in the right hand side of this equation have clear physical meaning – the adatom concentration decreases because of desorption and attachment of adatoms to the steps. On the other hand it increases because of detachment of atoms from the steps (see the third term). The transparency of the steps also leads to a decrease of the concentration due to direct jumps of adatoms to the neighboring terraces. Finally, the jumps of the adatoms from the neighboring terraces to the $i-th$ terrace lead to an increase of the concentration $n_i$ (see the last two terms).

Accounting for the dependence of the equilibrium concentration on the size of the neighbouring terraces [24, 25]

$$n_s^e(i) = n_s^e\left[1 + \tilde{A}\left(\frac{1}{l_{i-1}^3} - \frac{1}{l_i^3}\right)\right] \quad (14)$$

where

$$\tilde{A} = \frac{2\Omega A}{kT} \quad (15)$$

we can write the final form of the eq.(13) (here $A$ is the coefficient of step repulsion of elastic and entropic nature). In dimensionless variables $\tau = \frac{Kt}{l}$, $\eta_i = \frac{l_i}{l}$, $c_i = \frac{n_i}{n_s^e}$, $\varepsilon = \frac{\tilde{A}}{l^3}$, $P_k = \frac{P}{K}$ we obtain

$$\frac{dc_i}{d\tau} = -\frac{c_i}{\tau_s'} - \frac{2}{\eta_i} c_i + \frac{2}{\eta_i} + \frac{1}{\eta_i}\varepsilon\left(\frac{1}{\eta_{i-1}^3} - \frac{1}{\eta_{i+1}^3}\right) + \frac{P_k}{\eta_i}(c_{i+1} - 2c_i + c_{i-1}) + \frac{Fd_s}{kT\eta_i}(c_{i-1} - c_{i+1}) \quad (16)$$

The time evolution of the terraces is governed by the equations

$$\frac{dl_i}{dt} = -K\Omega\left[\begin{array}{l}\left[\left(n_{i+1}\left(1 - \frac{Fd_p}{kT}\right) - n_s^e(i+1)\right) + \left(n_i\left(1 + \frac{Fd_p}{kT}\right) - n_s^e(i+1)\right)\right] \\ -\left(n_{i-1}\left(1 + \frac{Fd_p}{kT}\right) - n_s^e(i)\right) - \left(n_i\left(1 - \frac{Fd_p}{kT}\right) - n_s^e(i)\right)\end{array}\right] \quad (17)$$

which immediately follow from the Eq.(2). In dimensionless variables the last set of equations reads

$$\frac{d\eta_i}{d\tau} = -n_s^e\Omega\left\{c_{i+1} - c_{i-1} + 2\varepsilon\left(\frac{1}{\eta_{i+1}^3} - \frac{2}{\eta_i^3} + \frac{1}{\eta_{i-1}^3}\right) - \frac{Fd_s}{kTP_k}(c_{i-1} - 2c_i + c_{i+1})\right\} \quad (18)$$

## 5. Linear stability analysis

The simplest solution of the equations (16) and (18) is $\eta_i = 1$ and $c_i = c_0 = \dfrac{1}{1+1/2\tau'_s}$ which is an equidistant step distribution and the corresponding average concentration of adatoms. To check the stability of this solution with respect to small fluctuations of the terrace size and adatom concentration we consider small deviations from $\eta_i = 1$ and $c_i = c_0$, i.e., $\eta_i = 1 + \Delta\eta_i(\tau)$ and $c_i = c_0 + \Delta c_i(\tau)$ (with the additional requirement $\Delta\eta_i(\tau) \ll 1$ and $\Delta c_i(\tau) \ll c_0$). Substituting these expressions into Eqs. (16) and (18) we make use of series expansion to obtain linear equations for $\Delta\eta_i(\tau)$ and $\Delta c_i(\tau)$.

Following the routine we look for a solution of the type

$$\Delta\eta_j = e^{ijq}\eta_q(\tau), \qquad \Delta c_j = e^{ijq+i\phi}c_q(\tau) \qquad (19)$$

where $q$ is a wave number and we already use $i$ to denote the imaginary unit and $j$ to denote the sequence number of the terrace. In addition we allow for a phase shift $\phi$ of the wave describing the adatom concentrations $\Delta c_j$ with respect to the wave describing the terrace widths $\Delta\eta_j$. In this way we arrive to

$$\frac{d\eta_q}{d\tau} = a_{11}\eta_q(\tau) + a_{12}c_q(\tau) \qquad (20)$$

$$\frac{dc_q}{d\tau} = a_{21}\eta_q(\tau) + a_{22}c_q(\tau)$$

Here, the coefficients are given by the following expressions

$$a_{11} = -12 n_s^e \Omega \varepsilon (1 - \cos q)$$

$$a_{12} = -2 e^{i\phi} n_s^e \Omega \left[ i \sin q + \frac{f}{P_k}(1 - \cos q) \right]$$

$$a_{21} = 2 e^{-i\phi} \left[ 3i\varepsilon \sin q - (1 - c_0) \right] \qquad (21)$$

$$a_{22} = -2\left[ (1 + 1/2\tau'_s) + P_k(1 - \cos q) + if \sin q \right]$$

where $f = \dfrac{F d_s}{kT}$.

The set of differential equations (20) has a solution of the type $e^{s\tau}$ where $s$ satisfies the algebraic equation

$$(a_{11} - s)(a_{22} - s) - a_{12} a_{21} = 0 \qquad (22)$$

The amplitude of the fluctuations in the terrace width distribution grows ($\eta_q \sim e^{s\tau}$) when the real part of $s$ is positive. To obtain a reasonably simple expression for the real part of

$$s_{1,2} = \frac{1}{2}\left[ (a_{11} + a_{22}) \pm \sqrt{(a_{11} + a_{22})^2 - 4(a_{11}a_{22} - a_{12}a_{21})} \right] \qquad (23)$$

we use appropriate approximations described in the Appendix A. As a result the real part of $s$ is given by (see eqs.(A4)-(A6))

$$s = B_2 q^2 - B_4 q^4 \qquad (24)$$

$$B_2 = 4 n_s^e \Omega \left[ f(1 - c_0) - 6\varepsilon \right]$$

$$B_4 = 4n_s^e \Omega P_k \left[ f(1-c_0) - \frac{9}{2}\varepsilon \right]$$

These expressions are valid for small values of the wave number $q \ll 1$. The term $(1-c_0) = \frac{n_s^e - n_s}{n_s^e}$ which multiplies the factor $f$ in the expressions for $B_2$ and $B_4$ is very essential since it can change the sing of $B_2$ and $B_4$. In other words the local deviation from equilibrium $\frac{n_s^e - n_s}{n_s^e}$ can have an impact on the stability of the vicinal surface. This circumstance needs some comments. In a presence of non-zero deposition rate $R$ an additional term, accounting for the increase of the adatom concentration, appears in the right hand side of eq. (16). This term reads $c_{st}/\tau_s'$ where $c_{st} = R\tau_s/n_s^e$. Obviously, when the deposition rate has its equilibrium value $R_e$ one obtains $c_{st} = R_e\tau_s/n_s^e = 1$. This means that $c_{st} < 1$ corresponds to undersaturation and $c_{st} > 1$ corresponds to supersaturation at the crystal surface. In a presence of deposition rate the steady state solution of the modified equation (16) is

$$c_0 = \frac{1 + c_{st}/2\tau_s'}{1 + 1/2\tau_s'} \tag{25}$$

and the term $(1-c_0)$ is positive in crystal sublimation and negative in crystal growth. That is why the eq.(24) predicts instability ($s > 0$) during sublimation at $f > 0$ provided the term (in the expression for $B_2$) containing $\varepsilon$ is smaller than the term containing $f$. In contrast, during crystal growth ($1-c_0 < 0$) instability of a train of transparent steps is expected at $f < 0$, i.e. at step down direction of the electromigration force.

When the coefficient $B_2$ is positive, i.e. $f(1-c_0) > 6\varepsilon$ (see eq.(24)) the step train is unstable for fluctuations with wave number $0 < q < \sqrt{\frac{B_2}{B_4}}$ and the most unstable mode has a wave number $q_{max} = \sqrt{\frac{B_2}{2B_4}}$. A very simple expression for $q_{max}$ is obtained in the case of strong instability, i.e., when the destabilizing term $f(1-c_0)$ is much larger than the stabilizing term $6\varepsilon$. Under this condition one gets

$$q_{max} = 1/\sqrt{2P_k} \tag{26}$$

and therefore the most unstable mode has a wave length (expressed as a number of terraces)

$$\lambda_{max} = 2\pi\sqrt{2P_k} \tag{27}$$

As seen the wave length of the most unstable mode depends only on the relative transparency $P_k = P/K$.

It is of interest to approach the problem of the most unstable mode without using series expansion of the expression (A2). The amplification rate $s_r(q)$ displays a maximum at the wave number $q_{max}$ which satisfies the equation $\frac{ds_r}{dq}=0$. Making use of the inequality $1/P_k \ll 1$ we obtain (see Appendix B)

$$\lambda_{max} = \pi\sqrt{2P_k} \tag{28}$$

which is valid when $\left|\frac{\varepsilon}{f(1-c_0)}\right| \ll 1$.

As seen more accurate treatment gives 2 times smaller value for the wave length of the most unstable mode.

Finally, it is instructive to rewrite the instability condition $B_2 > 0$ in terms of a drift velocity of the adatoms $V_{drift} = \frac{FD_s}{kT}$. Having in mind that $f = \frac{Fd_s}{kT}$ and $d_s = \frac{D_s}{K}$ one can write

$$B_2 = \frac{4n_s^e \Omega}{K}\left[V_{drift}\frac{n_s^e - n_s}{n_s^e} - V_{cr}\right] \tag{29}$$

where

$$V_{cr} = \frac{12K\Omega A}{kTl^3} \tag{30}$$

is exactly the same critical velocity as $V_{cr}$ obtained in a solution of a problem with a rather different physics (fast sublimation or growth in absence of any destabilizing factors) [18].

It is interesting to compare the results of the linear stability analysis of the kinetic equations describing our discrete model with the results obtained in Ref.[15]. Sato et al performed a linear stability analysis of the continuum model of a vicinal surface with very transparent steps (this model was previously formulated by Stoyanov [13]). Their central result for the case of sublimation at a zero deposition rate reads (in our notation)

$$B_2 = \frac{n_s^e \Omega}{2\tau_s'}[f - 6\varepsilon] \tag{31}$$

which is somewhat different from the expression (24) we obtained for the discrete model of vicinal surface.

## 6. Non-linear dynamics of transparent steps

Essential element of the derivation of the equations (16) is the use of the expression (7) for the electromigration induced gradient of the adatom concentration on the terraces. Since the expression (7) was obtained for equidistant configuration of the steps, the equations (16) are, strictly speaking, not applicable in the late stages of vicinal surface instability where bunches of steps are separated by large terraces. Nevertheless, we shall study the non-linear dynamics of steps by numerically integrating the equations (16) and (18). This procedure is justified only in the early stages of step bunching instability and it gives information for the wave length of the most unstable mode. As for the late stages of the non-linear dynamics we only hope to get some qualitative idea about the time evolution of the instability of a vicinal surface with transparent steps. The quantitative results, however, are not expected to be reliable since the real gradient of the adatom concentration on a given terrace could differ significantly from the value predicted by eq.(7).

For the integration of eqs.(16) and (18) we need numerical values of the parameters $\tau_s^/ = \frac{K\tau_s}{l}$, $\varepsilon = \frac{\tilde{A}}{l^3}$, $P_k = \frac{P}{K}$, $f = \frac{Fd_s}{kT}$ and $n_s^e \Omega$. These values should be in agreement with the assumptions already made in this model. For instance, the characteristic lengths are assumed to satisfy the inequalities $d_s \geq l$ and $d_p \ll l$ in order to reduce the expression (6) to $grad(n_s) \approx n_s^e \frac{2Fd_p}{kT}$. These inequalities, however, immediately show that $\frac{d_s}{d_p} = \frac{P}{K} = P_k > 10$ (it is reasonable to remind that Tanaka et al [21] estimated from their experimental results $P_k = 36$ for the steps at Si(001) surface). In the numerical integration of the eqs.(16) and (18) we used values of the parameter $P_k$ between 10 and 50.

The values of the parameter $\tau_s^/ = \frac{\tau_s K}{l}$ are restricted by the requirement $l^2 \ll \lambda_s^2 = D_s \tau_s = \frac{D_s}{K}\tau_s K = d_s l \frac{\tau_s K}{l} = d_s l \tau_s^/$ which is equivalent to $\frac{l}{d_s} \ll \tau_s^/$. As mentioned above the characteristic length $d_s$ is assumed to be $d_s \geq l$ which immediately means $\tau_s^/ \gg 1$. The parameter $\tau_s^/$ determines the deviation of the real concentration of adatoms during sublimation from the equilibrium concentration since $c_0 = \frac{n_s}{n_s^e} = \frac{2\tau_s^/}{1+2\tau_s^/}$.

We shall study the step dynamics at $\tau_s^/ > 5$ (which means $n_s^e \geq n_s \geq 0.9 n_s^e$, i.e., near to equilibrium). For the case of growth, however, we can go well away from the equilibrium by increasing the deposition rate while keeping the value $\tau_s^/ > 5$. The parameter $\varepsilon$ describes the step-step repulsion and values in the interval $10^{-8} < \varepsilon < 10^{-3}$ will be used in the numerical integration of the equations (the large interval of values of $\varepsilon = \frac{\tilde{A}}{l^3}$ is used to see the impact of the surface orientation on the instability).

The parameter $f = \frac{Fd_s}{kT}$ is estimated on the basis of the experimental results of Fujita et al [3] for the shape of the step bunches. This interpretation [3] provides the value $\frac{F}{A} = 2 \times 10^{-6}$ [nm$^{-2}$] at $1250°C$. Making use of the value $A = 0.1$ [eVnm] and assuming $d_s = \frac{D_s}{K} = 10 l$ we estimate $f \approx 0.6 \times 10^{-3}$ for a temperature $1230°C$ and a vicinal surface with a terrace width $l = 100$ [nm]. As will be seen below the main difficulties in reproducing the experimental observations are related to the values of the parameter $f$ - step bunching occurs at unrealistically high values of the electromigration force $F$. That is why we assumed $d_s = 10l$ in the estimation of $f$. This assumption, however, automatically means $\frac{d_s}{d_p} = \frac{P}{K} = P_k > 100$. Finally, we shall explore the step dynamics at $n_s^e \Omega = 0.1$ and $n_s^e \Omega = 0.01$ in agreement with the experimental findings [26]. Smaller values of this parameter lead to very slow evolution of the terrace distribution and very long computation time.

The numerical integration of the eqs.(16) and (18) for $\tau_s' = 5$, $P_k = 20$, $n_s^e \Omega = 0.1$, $\varepsilon = 10^{-7}$ manifests step bunching at $f = 5 \times 10^{-3}$ which is one order of magnitude larger than the value estimated above.

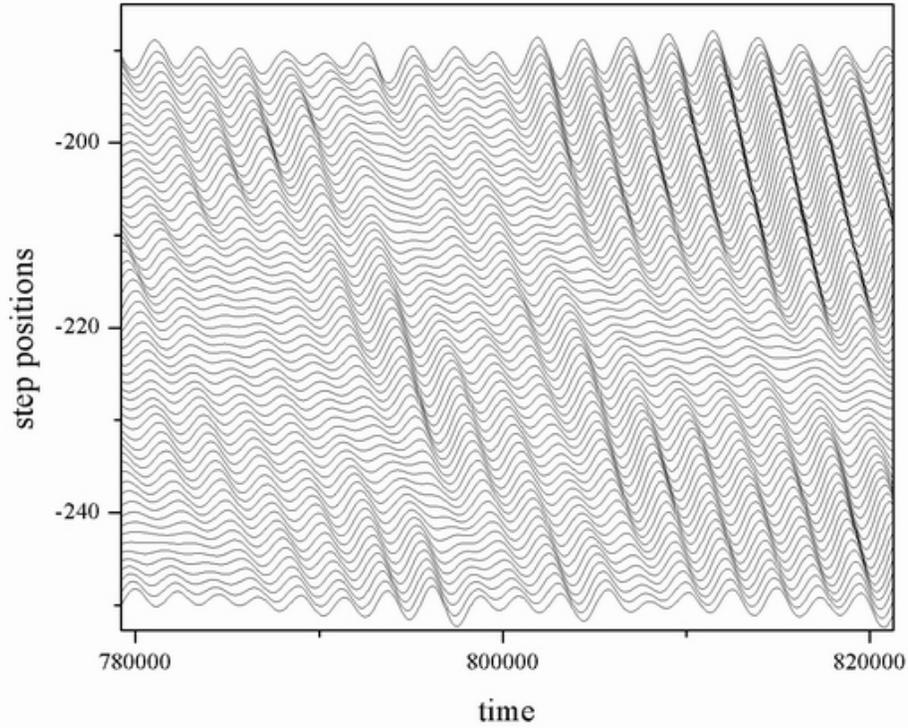

**FIG.2.** The step trajectories during sublimation given in a frame moving with the average velocity of the steps. Model parameters are:
$n_s^e \Omega = 0.1; f = 5 \times 10^{-3}; \varepsilon = 10^{-7}; P_k = 20; \tau_s' = 5$

As seen in Fig.2 step dynamics manifests formation of relatively small bunches which disappear at some moment but soon after that reappear in a place close to the place of disappearance.

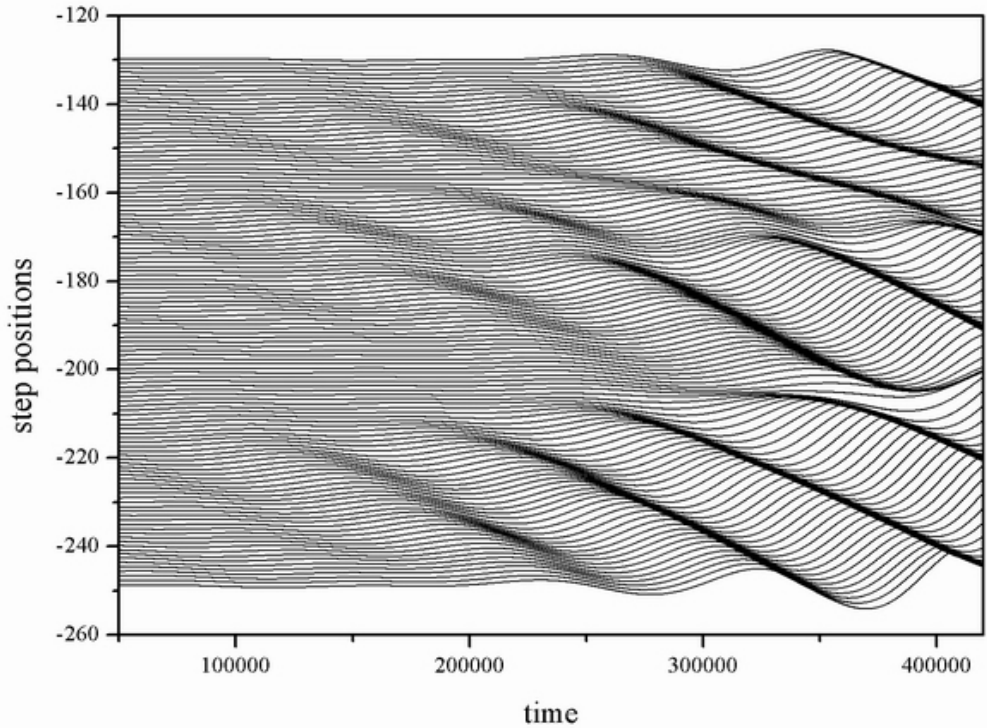

**FIG.3.** Step trajectories at larger electromigration force. Model parameters are:
$n_s^e \Omega = 0.01; f = 1; \varepsilon = 10^{-5}; P_k = 20; \tau_s' = 5$

To obtain larger bunches we had to increase the electromigration parameter to an unrealistically high value $f = 1$ (see Fig.3). Further increase $f = 3$ produced the step trajectories shown in Fig.4.

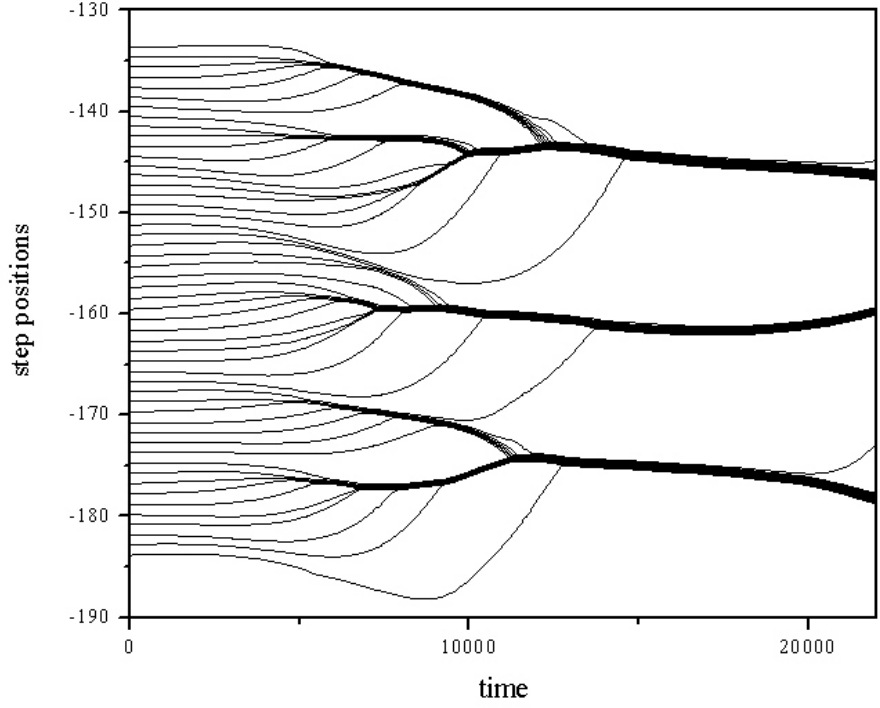

**FIG.4.** Step trajectories at model parameters:
$n_s^e \Omega = 0.01; f = 3; \varepsilon = 10^{-4}; P_k = 5; \tau_s' = 2$

As discussed in the beginning of this section the results obtained by numerical integration give only some idea about the late stages of the step bunching instability. The step bunches in Fig.3 and 4 are strongly compressed (the minimum interstep distance is $l_{min} = 0.04l$ in Fig.3 and $l_{min} = 0.025l$ in Fig.4). This is probably a result of the unrealistically high values of the electromigration force (in Fig.2 we have $l_{min} = 0.3l$) and, also the fact that the eqs.(16) and (18) do not describe accurately the late stages of the step bunching instability (simultaneous existence of neighbouring terraces of very different size).

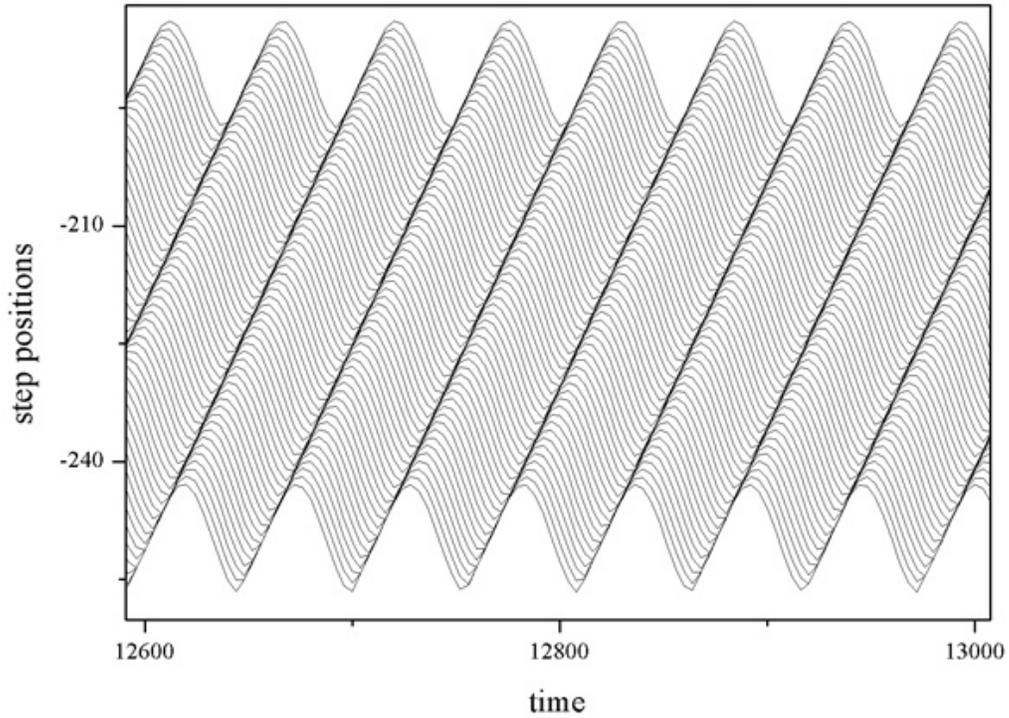

**FIG.5.** Step trajectories during growth in an absence of electromigration force. Model parameters are: $n_s^e \Omega = 0.1; f = 0; \varepsilon = 10^{-5}; P_k = 20; \tau_s' = 5; c_{st} = 50$

In the case of growth in an absence of electromigration we observed propagation of step density waves (see Fig.5) similar to the dynamics of non-transparent steps [18].

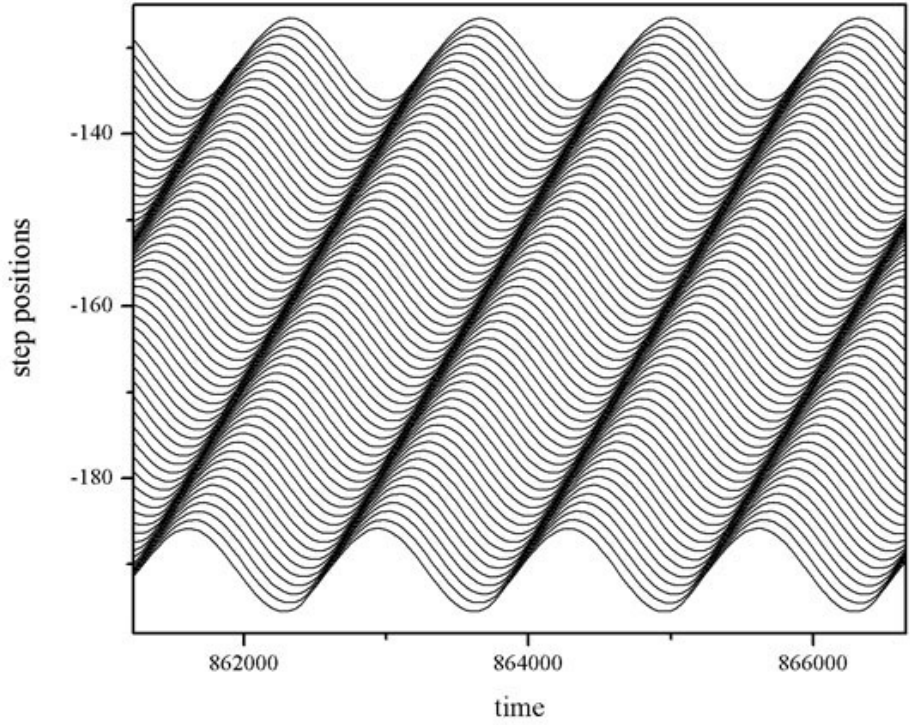

**FIG.6.** Growth in a presence of electromigration force. Model parameters:
$n_s^e \Omega = 0.1; f = -5 \times 10^{-2}; \varepsilon = 10^{-3}; P_k = 20; \tau_s' = 10; c_{st} = 5$

Larger waves propagate in a presence of a force with step-down direction (see Fig.6).

### 7. Concluding remarks

We believe it is essential to account for the non-steady state effects when treating a non-local electromigration model for sublimation and growth of crystal surfaces with transparent steps. The reason is that one atom, detached from a given step, can cross 10 or 20 steps before attaching to a kink position. Such a long trip requires considerable time. In other words, what is happening at a given step in a given moment depends on the "history" of a considerable part of the crystal surface around this step. This is the basic idea of our approach. The considerations are restricted to the case of slow kinetics at the steps and fast diffusion on the terraces.

The central results of this study are:

1. The vicinal surface with transparent steps is unstable when the drift velocity $V_{drift} = \dfrac{FD_s}{kT}$ of the adatoms multiplied with the relative deviation from equilibrium $(1-c_0) = \dfrac{n_s^e - n_s}{n_s^e}$ is larger than the critical velocity of the steps $V_{cr} = \dfrac{12K\Omega A}{kTl^3}$. It is important to note that instability takes place at a positive value of the product $\frac{n_s^e - n_s}{n_s^e} V_{drift}$. Since $\frac{n_s^e - n_s}{n_s^e}$ is positive during sublimation and negative during growth step bunching takes place at a positive drift velocity in the first case and at a negative $V_{drift}$ (step-down direction of the electromigration) in the second case.

2. The most unstable mode has a wave length (expressed as a number of terraces) given by the expression $\lambda_{max} = \pi\sqrt{2P_k}$. In other words, the wave length (as a number of terraces) is independent of the surface orientation as well as on the magnitude of step repulsion and the electromigration force. The quantity $\lambda_{max}$ depends only on the relative transparency of the steps $P_k = P/K$.

3. The numerical integration of the kinetic equations of our model manifests strong bunching of the transparent steps when the product $V_{drift} \frac{n_s^e - n_s}{n_s^e}$ is much larger than $V_{cr}$. In an absence of electromigration the numerical results manifest a formation of step density waves (see Fig.5). When the electromigration force has a step-down direction instability occurs during growth. The bunches are relatively large and they form a periodic array (see Fig.6).

The confrontation with the reported [3-5, 14, 27,28] experimental observations reveals several strong and several weak points of our model. The linear stability analysis and the numerical integration of eqs.(16) and (18) predicts step bunching instability at step-up electromigration of the adatoms during sublimation whereas step-down adatom drift induces instability during growth. These predictions are in good agreement with the experimental findings [4, 14]. Our model, however, cannot reproduce the observed [4] step bunching during vapour-crystal equilibrium. Here one should mention the experiments of Peltz et al [27] who claim the step bunching in the second temperature interval $(1040°C - 1190°C)$ to occur at step-up direction of the electric current during sublimation, equilibrium and growth of (111) Si crystal face. These findings contradict our results for equilibrium and growth.

As far as the wave length of the most unstable mode is concerned our formula (28) is in contradiction with the reported [28] unique wavelength $\lambda = 4\mu m$. Our result $\lambda_{max} = \pi\sqrt{2P_k}$ should be multiplied by $l$ in order to obtain the wavelength $\lambda$ discussed in Ref.[28]. Calculated in this way $\lambda$ increases with $l$ in clear contradiction with the unique wavelength reported by Leroy et al [28] for samples with variation of $l$ by two orders of magnitude.

Another weak point of the model is related to the unrealistically high values of the electromigration force necessary to obtain bunching with coarsening (see Figs.3 and 4) by numerically integrating eqs.(16) and (18). The reason for this circumstance is probably related to the fact that these equations do not provide an accurate description of the step dynamics in the late stages of the step bunching instability.

In conclusion, the non-local electromigration destabilizes the vicinal surface when the expression $V_{drift}(n_s^e - n_s)/n_s^e$ is positive and larger than $V_{cr}$. This step bunching instability, however, manifests some features which contradict the reported experimental observations.

**Appendix A**

The square root in eq. (19) can be rewritten in the form

$$\sqrt{(a_{11} + a_{22})^2 - 4(a_{11}a_{22} - a_{12}a_{21})} = \sqrt{(a_{11} + a_{22})^2} \sqrt{1 - 4\frac{(a_{11}a_{22} - a_{12}a_{21})}{(a_{11} + a_{22})^2}} \quad (A1)$$

Further we find expressions for the real and the imaginary part of the ratio $\frac{(a_{11}a_{22} - a_{12}a_{21})}{(a_{11} + a_{22})^2}$ by making use of appropriate approximations. The real and the imaginary part of the square root $\sqrt{1 - 4\frac{(a_{11}a_{22} - a_{12}a_{21})}{(a_{11} + a_{22})^2}}$ are handled in the way described in [18, 19]. Finally, we obtain a rather complicated expression for the real part of $s$

$$s_r(q) = \frac{4n_s^e \Omega q^2}{F_1(q^2)} \left\{ (1 - c_0)f \left[ \frac{1}{P_k} + \frac{4}{F_1(q^2)} \right] - 3\varepsilon \left[ F_2(q^2) + \frac{8}{F_1(q^2)} \right] \right\} \quad (A2)$$

where
$$F_1(q^2) = 1/\tau_s' + 2 + P_k q^2$$
$$F_2(q^2) = 1/\tau_s' + P_k q^2 \quad (A3)$$

Expanding this expression in power series of $x = q^2$ and neglecting the terms with $n > 2$ we arrive to
$$s_r = B_2 q^2 - B_4 q^4 \quad (A4)$$

Having in mind that the relative transparency $P_k$ and the dimensionless life time of the adatoms $\tau_s'$ should satisfy the inequalities $P_k > 10$ and $\tau_s' \geq 5$ we neglect the terms $1/P_k$ and $1/\tau_s'$ with respect to 2 and obtain
$$B_2 = 4n_s^e \Omega [f(1 - c_0) - 6\varepsilon] \quad (A5)$$

$$B_4 = 4n_s^e \Omega P_k \left[ f(1 - c_0) - \frac{9}{2}\varepsilon \right] \quad (A6)$$

**Appendix B**

It is of interest to approach the problem of the most unstable mode without using series expansion of the expression (A2). The fluctuation amplification rate $s_r(q)$ displays a maximum at the wave number $q_{max}$ which satisfies the equation $\frac{ds_r}{dq} = 0$. Taking a derivative of the expression (A2) and simplifying the result by making use of the inequalities $1/P_k \ll 1$ and $1/\tau_s' \ll 2$ we arrive to the following cubic equation for the variable $y = P_k q^2$

$$3\varepsilon y^3 + 18\varepsilon y^2 + 4f(1-c_0)y - 8[f(1-c_0) - 6\varepsilon] = 0 \qquad (B1)$$

This equation has a simple root $y_1 \approx 2$ when the inequality $\dfrac{\varepsilon}{(1-c_0)f} \ll 1$ is satisfied. Having in mind the definition of this variable ($y = P_k q^2$) one obtains

$$q_{max} = \sqrt{y_1/P_k} \approx \sqrt{2/P_k} \qquad (B2)$$